\documentclass[11pt,a4paper,dvipdfmx]{article}
\usepackage{aer}
\usepackage{aer-osborn}
\usepackage{graphicx}
\usepackage[bf,hang,margin=7truemm]{caption}[2007/04/09]
\usepackage{amsmath}
\usepackage[]{natbib}
\def\figp#1#2#3#4{
\begin{figure}[!tp]
\begin{center}
\includegraphics[width=#3\textwidth,bb=#4]{#1.pdf}
\caption{#2}
\label{fig:#1}
\end{center}
\end{figure}}
\begin{document}
% Pareto indices
\def\muf{{\mu_{\rm F}}}
\def\muw{{\mu_{\rm W}}}
\def\mus{{\mu_{\rm S}}}
% PDF and Cumulative Probability
\def\ps{p^{(\rm S)}(c)}
\def\psc{P^{(\rm S)}_>(c)}
\def\pf{p^{(\rm F)}(c)}
\def\pfc{P^{(\rm F)}_>(c)}
\def\pw{p^{(\rm W)}(c)}
\def\pwc{P^{(\rm W)}_>(c)}
\def\pfd{p^{(\rm F)}(0)}
\def\pfe{P^{(\rm F)\prime}(0)}
% Misc. for equilibrium discussion
%\def\wn{w_{\{n\}}}
%\def\pn{P_{\{n\}}}
\def\sumk{\sum_{k=1}^K}
\def\pk{p_k}
%\def\dtil{\widetilde{D}}
%\def\dc{\Delta_c}
% Averages of productivity
\def\act{\langle c \rangle_{\beta}}
\def\actn#1{\langle c^#1 \rangle_{\beta}}
\def\ac{\langle c \rangle_{0}}
\def\acn#1{\langle c^#1 \rangle_{0}}
\def\pc{P(c)}
\def\pb{p^{\rm (GB2)}(c)}
\def\xx{\biggl(\frac{c}{c_1}\biggr)}
\def\xxi{\biggl(\frac{c_1}{c}\biggr)}
%% End of Aoyama's defs
\def\addnum#1{$^#1$}
\title{\bf Superstatistics of Labour Productivity 
in Manufacturing and Nonmanufacturing Sectors}
\author{\sl 
Hideaki Aoyama \addnum1, 
Yoshi Fujiwara \addnum2,
Yuichi Ikeda \addnum3,
\\[5pt]\sl
Hiroshi Iyetomi \addnum4,
and 
Wataru Souma \addnum2}
\date{}
\maketitle

%\vspace{-72mm}
%\includegraphics[width=0.4\textwidth,bb=0 0 510 167]{omoshiroki.jpg}
%\vspace{40mm}

\begin{center}
{\small\sl
\addnum1 
Department of Physics, Kyoto University, Kyoto 606-8501, Japan\\[3pt]
\addnum2 
NiCT/ATR CIS, Applied Network Science Lab., Kyoto 619-0288, Japan\\[3pt]
\addnum3 
Hitachi Research Institute, 4-14-1 Soto-kanda, Chiyoda-ku, Tokyo, 101-8010,
Japan\\[3pt]
\addnum4
Department of Physics, Niigata University, Niigata 950-2181, Japan
}

\end{center}

\abstract{
Labour productivity distribution (dispersion) is studied both theoretically
and empirically.
Superstatistics is presented as a natural theoretical framework for productivity.
The demand index $\kappa$ 
is proposed within this framework as a new business index.
Japanese productivity data covering 
small-to-medium to large firms from 1996 to 2006
is analyzed and the power-law for both firms and workers is established.
The demand index $\kappa$ is evaluated in the manufacturing sector.
A new discovery is reported for the nonmanufacturing (service) 
sector, which calls for expansion of the superstatistics framework 
to negative temperature range.
}\\[20pt]
% JEL numbers below
% Keywords
% Correspondence
{\bf Correspondence:}
Hideaki Aoyama, Physics Department, Kyoto University,
Kyoto 606-8501, Japan. Email: hideaki.aoyama@scphys.kyoto-u.ac.jp\\[10pt]
% revision history
% Original was on Dec. 11, 2008.
Revised on Jan.~12th, 2009.\\[10pt]
% Acknowledgements
\vfill
\hrule
\vspace{6pt}
\noindent{\sl\footnotesize $^*$
The authors would like to thank Professor Hiroshi Yoshikawa
for previous collaborative work and Professor Masanao Aoki for encouraging
us at various stages of our research in econophysics over the years. 
We would also like to thank the CRD association and its chairman, Mr.\ Shigeru Hikuma,
for his help and advise in using their database.
Computing Facility at the Yukawa Institute for Theoretical Physics was 
used for part of the numerical computation.}
\thispagestyle{empty}
%%%%%%%%%%%%%%%%%%%%%%%%%%%%%%%%%%%%%%%%%%%%%%%%%%%%%%%%%%%%%%%
\newpage
\section{Introduction}
Standard equilibrium theory in economics implies that
the labour productivity is equal among firms and sectors.
This was shown to be wrong by a detailed study of the real data \citep{ayif1}.
Facing this situation, one may argue  that 
(i) since the real data is slightly
different from ideal case that the economic theory deals with
and is contaminated with various inaccuracies and errors, so slight 
deviation from the theoretical prediction is unavoidable and is even expected,
and besides,
(ii) since the equilibrium theory is self-consistent, reasonable and convincing
it must be true,
These claims are not valid, as (i) the distribution
of productivity is wide-spread; it is not even a normal distribution
or log-normal distribution as expected from contamination argument, 
but does obey Pareto law (power law) \citep{pareto} for large productivity, that is,
the distribution has the distinct characteristics of fat tails, and 
(ii) there may exist other theories that are far more convincing and
the validity of the theory can be judged only by facing the true nature
of the subject.
Indeed, physics, or any other discipline of exact science managed to
develop to the current status just by following (ii):
No matter how the pre-Copernicus theory is reasonable, beautiful and convincing,
earth moves; no matter how the idea of absolute time in the Newtonian mechanics, 
which by the way underlies the current equilibrium theory of economics,
seems unavoidable,
Einstein's relativity theory describes the true nature of time and space.
These and other numerous historical examples in exact science teaches us that
we need to face the phenomena seriously and has to construct
the theory that meets its demand. 
Simply put, we need to take scientific approach.

Such was the thought behind the study of the productivity
by \citet{ayif1}, who
proposed the superstatistics theory in statistical physics
as the theoretical framework for the productivity.

In this paper, we further advance the superstatistics theory of
productivity by examining the whole spectrum of firms
in Japan, while in the previous work of \citet{ayif1} and
\citet{ayif2} the data was limited to listed firms.
Furthermore, we analyse the manufacturing and nonmanufacturing
sectors separately.
In Section II (and in the Appendix), we present the superstatistics framework for the
productivity for completeness. Then in Section III, after explaining
the nature of the database and the method of analysis, we
present the results for the manufacturing sector and the
nonmanufacturing sector separately. We also study the distribution
of the productivity of business sectors.
Section IV contains some conclusions and discussions on the necessities
of extending the superstatistics framework.

\section{Superstatistics theory of productivity}
We first review the superstatistics theoretical framework of 
productivity proposed by \citet{ayif1,ayif2} in a concise manner.
The data analysis and the discussion of the evaluation of
the property of the aggregate demand is done in later sections
using this framework.

\subsection{Statistics}
Yoshikawa and Aoki \citep{yoshikawa, aokiyoshikawa} proposed 
an equilibrium theory of productivity distribution several years ago. 
Its essence is the equilibrium theory statistical physics, where
the most common distribution is realized under given constraint.
Its beauty lies in the fact that it does not depend the
details of the individual properties and interactions among 
constituents (firms in economics and atoms and molecules in 
statistical physics). Let us first review it very briefly.

We label firms by an index $k=1,2,\dots,K$, where $K$ is the total number of firms,
the number of workers at the $k$-th firm by $n_k$,
the productivity of the $k$-th firm by $c_k$, all for a given year.
There are two constraints on these quantities.
\begin{description}
\item[(i) Total number of workers $N$:]
\begin{equation}
\sumk n_k = N.
\label{ncons}
\end{equation}
\item[(ii) The aggregate demand $\tilde{D}$:]
The sum of firm's production is the total production, which is equal to the aggregate 
demand $\tilde{D}$;
\begin{equation}
\sumk n_k c_k =\tilde{D}.
\label{dcons}
\end{equation}
Implicit here is that we are dealing with the {\it mean} labour productivity
\begin{equation}
c:=\frac{Y}{L}\, ,
\label{yoverl}
\end{equation}
where $Y$ is the value added and $L$ is the labour (in number of workers).
Although this is different from the
{\it marginal} labour productivity 
$c^{\rm (marginal)}:=\partial Y /\partial L$ 
relevant in the standard equilibrium theory, this difference is irrelevant
as we will elaborate later.
\end{description}

By using the standard proposition that the distribution 
that maximizes the probability under these constraint is realized in nature, 
which is equivalent to the entropy-maximization, the Boltzmann law is obtained.
This states that the probability $\pk$ of the worker's productivity being equal to $c_k$
is the following.
\begin{equation}
\pk:=\frac{\langle n_k\rangle}{N}=\frac1{Z(\beta)} \, e^{-\beta c_k},
\label{boltz}
\end{equation}
where $Z(\beta)$ is the usual partition function:
\begin{equation}
Z(\beta):= \sumk e^{-\beta c_k}.
\label{partid}
\end{equation}
This guarantees the normalization of the probability $\pk$;
\begin{equation}
\sumk\pk=1.
\end{equation}
The parameter $\beta$ is inverse-temperature determined by the mean demand $D$ as follows:
\begin{equation}
D:=\frac{\tilde{D}}{N}=-\frac{d}{d\beta}\ln Z(\beta).
\label{dnbeta}
\end{equation}

In our database, we have nearly half a million firms and
about 10 million workers (see Fig.\ref{fig:nfirmsworkers}).
Therefore, it is most appropriate to use the continuous notation, 
in which the probability distribution function (pdf) of the firm's productivity 
is denoted by $\pf$, and the pdf of the worker's productivity by $\pw$.
From Eq.(\ref{boltz}), they satisfy the following:
\begin{equation}
\pw=\frac1{Z(\beta)} \, e^{-\beta c} \pf,
\label{pee}
\end{equation}
where the partition function is 
\begin{equation}
Z(\beta):=\int_o^\infty e^{-\beta c} \pf dc.
\label{parti}
\end{equation}

\subsection{Superstatististics}
Although the theoretical prediction (\ref{pee}) is both elegant and powerful, 
it is quite limited in the sense that it is realized in a stationary environment.
Namely, the demand (and thus the temperature) has to be constant.
However, the demand is rarely constant. Rather it is one of the most quickly
changing parameter \citep{yoshikawa}. Therefore, we need to 
expand the horizon of the theory to meet the changing environment.
Just such a theory, named  \textit{superstatistics} (statistics of statistics)
has been proposed recently by \citet{beckcohen} in the context of statistical physics.
In this theory, the system goes through changing external influences, but 
is in equilibrium described by Boltzmann distribution (\ref{pee})
within certain limited scale in time and/or space.
Therefore, the whole system can be described by an average over the Boltzmann factors, with
the weight given by the relative scales (in time and space) of the temperature ($1/\beta$),
which the system experiences.

This superstatistics was successfully applied to various systems \citep{beck1,beck2008}.
Most analogous to our economic system of firms and workers 
maybe the Brownian motion of a particle going thorough 
changing temperature and viscosity \citep{ausloos2006bph,lz}.
Our workers are the particles in Brownian motion: 
They move from firms to firms, which keeps trying to meet ever-changing demand
by employing and dismissing workers.
Therefore, the superstatistics is the right framework to deal with the
distribution of the workers.
The weighted average over the temperature
now replaces the Boltzmann factor $e^{-\beta  c}$:
\begin{equation}
B(c)=\int_0^\infty e^{-\beta c} f_\beta(\beta) d\beta.
\label{bc}
\end{equation}
In this equation the changing environment
is represented by the weight factor $f_\beta(\beta)$,
which is, in turn, is a function of the mean demand $D$ by Eq.(\ref{dnbeta}).
The pdf of worker's productivity (\ref{pee}) is now modified to;
\begin{equation}
\pw = \frac1{Z_B} \pf B(c).
\label{peess}
\end{equation}
The new partition function $Z_B$ in the above is given by,
\begin{equation}
Z_B=\int_0^\infty \pf B(c) dc.
\label{zbpss}
\end{equation}

Let us study what the superstatistics theory tell us for the high productivity region.
We concentrate on this region because,
as we will see by the data analysis in the next section,
both the firm's productivity and worker's productivity
obeys the Pareto's law (power law):
\begin{align}
\pf &\propto c^{-\muf-1},
\label{pfasym}\\
\pw &\propto c^{-\muw-1}.
\label{mue}
\end{align}
This feature brings advantage to the high-productivity study because
of the following reasons:
\begin{enumerate}
\item[(i)]
This feature is quite evident in the data and the Pareto indices $\mu_{\rm F, W}$
can be estimated reliably. 
In comparison, medium to low range sometimes shows two-peak
structure, which makes it difficult to extract notable, representative features.
(Elsewhere in this volume, \citet{soumaikeda} elaborates on this point.)
\item[(ii)]
As was proven by \citet{ayif1}, if the Pareto law holds for
the ``mean" productivity $c$, the same law with the same value of the
Pareto index holds for the marginal productivity $c^{\rm (marginal)}$
under a wide assumption.
\end{enumerate}

Let us now study the behaviour of Eq.(\ref{bc}) for large $c$.
This integration is dominated by the small $\beta$ region.
Thus the behaviour of the pdf $f_\beta(\beta)$ for small $\beta$ is critical.
Let us assume the following in this range:
\begin{equation}
f_\beta(\beta) \propto \beta^{-\gamma} \quad (\gamma<1),
\label{fbeta}
\end{equation}
where the constraint for the parameter $\gamma$ 
comes from the convergence of the integration in Eq.(\ref{bc}).
This leads to the following for large $c$:
\begin{equation}
B(c) \propto   \Gamma(1-\gamma)\, c^{\gamma-1}.
\label{bcgamma}
\end{equation}
Substituting this and the Pareto laws Eqs.~(\ref{pfasym}) and (\ref{mue}) 
into Eq.~(\ref{peess}),
we obtain the following:
\begin{equation}
\muw=\muf-\gamma+1.
\label{muef1}
\end{equation}

We note here that because of the constraint $\gamma <1$, this leads to 
the inequality
\begin{equation}
\muw > \muf.
\label{muwfineq}
\end{equation}
This becomes a critical test of this superstatistics theory of productivity,
which we come back to in the following section.

% modification below
The above derivation of Eq.(\ref{muef1})
proves, in effect, that the Pareto law for firms and that for workers
are compatible only if the temperature distribution obeys
Eq.~(\ref{fbeta}), as no other behaviour could result in the
power law (\ref{bcgamma}). In this sense, we see that
empirical observation leads to Eq.~({fbeta}).
This in turn leads to empirical laws to the
distribution and fluctuation of the demand through Eq.~(\ref{dnbeta}):
In this manner, the parameter $\gamma$ in the distribution of $\beta$ is related to
a parameter in distribution of $D$, which we denote by $f_{D}(D)$ by
the following relation:
\begin{equation}
f_\beta (\beta)d\beta=f_{D}(D)dD,
\end{equation}
Mathematical relation between $\beta$ and $D$ is studied in detail
in the appendix. Using the result (\ref{hight})
and Eq.(\ref{fbeta}), we find that 
\begin{equation}
f_{D}(D)\propto\left(\ac-D \right)^{-\delta},
\label{deltadef}
\end{equation}
with 
\begin{equation}
\gamma-1=
\begin{cases}
\delta-1 
&\mbox{for }2<\muf;\\
(\muf-1)(\delta-1)
&\mbox{for }1<\muf<2.
\end{cases}
\label{gammarho}
\end{equation}
Note that $f_{D}(D)$ has an upper limit $\ac$:
From Eq.(\ref{pee}), it is evident that as the temperature $1/\beta$ goes up,
workers move to firms with higher productivity.
As the temperature becomes infinity $\beta=0$, all the firms has
the same number of workers. Thus the total demand is
limited by the values achieved at this point, where $D=\ac$.

\figp{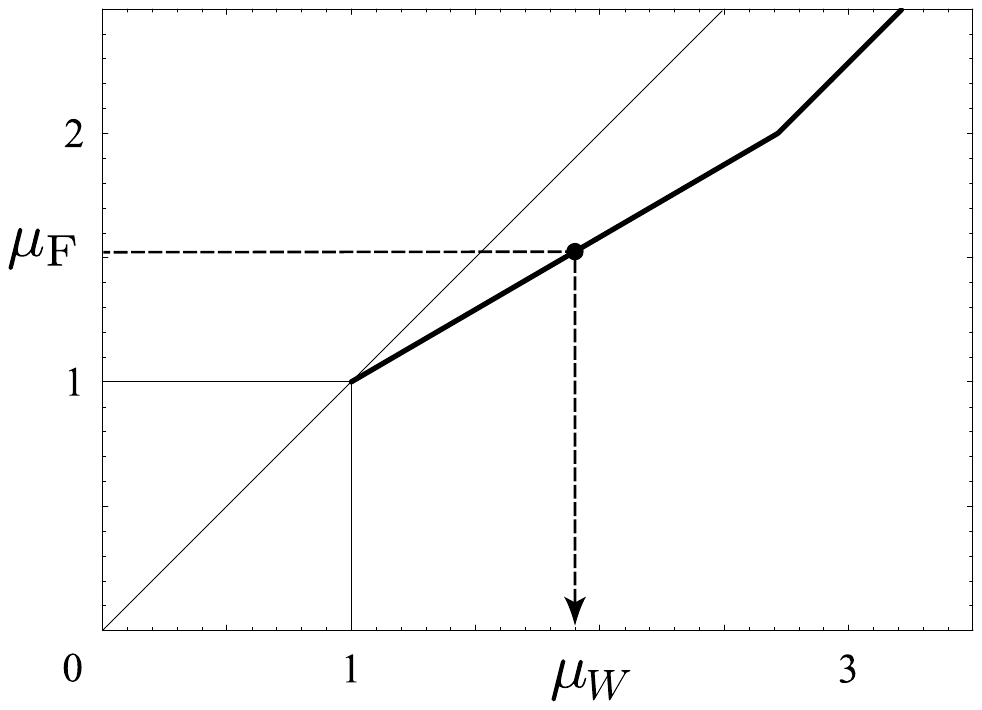}{Illustration of the relation between $\muw$ and $\muf$ (\ref{mumu}). 
The solid line is the relation (\ref{mumu}), and the filled circle 
is the data.}{0.4}{138 413 415 613}

Combining Eqs.(\ref{muef1}) and (\ref{gammarho}),
we reach the following relation between the Pareto indices:
\begin{equation}
\muw= \begin{cases}
\muf-\delta+1 & \mbox{for } 2<\muf \\
(\muf-1)(-\delta+1)+\muf &\mbox{for } 1<\muf<2.
\end{cases}
\label{mumu}
\end{equation}
This relation between $\mu _W$ and $\mu _F$ is illustrated in Fig.\ref{fig:mumu}.
The range of the parameter $\delta$ is $-\infty<\delta<1$
from the normalizability of the distribution of $f_{D}(D)$.
The upper limit may also be obtained from the constraint $\gamma<1$ 
and Eq.(\ref{gammarho}).
Because of this, Eq.(\ref{mumu}) predicts that $\muw$ is larger than $\muf$.
Also, Eq.(\ref{mumu}) has a fixed point at $(\muw,\muf)=(1,1)$;
the line defined by Eq.(\ref{mumu}) always passes through this point irrespective of
the value of $\delta$. 
The Pareto index for firms is smaller than that for workers, but it
cannot be less than one, because of the existence of this fixed point.

This way, the parameter $\delta$ calculated from $\muf$ and $\muw$
represents the behaviour of the demand close to its upper limit.
As a parameter with the same function, we propose the 
following parameter, which we call {\it Demand Index}:
\begin{equation}
\kappa:= \frac1{2-\delta}\ .
\end{equation}
This parameter is a monotonically increasing function of 
$\delta$ and ranges from 0 to 1. The limited range of $\kappa$ makes easy to handle and plot.
If $\kappa$ is close to one, the demand fluctuated to the high region significantly;
if it is equal to zero, the demand does not go very high (it could be 
dumping faster than any power law toward the upper limit).
\figp{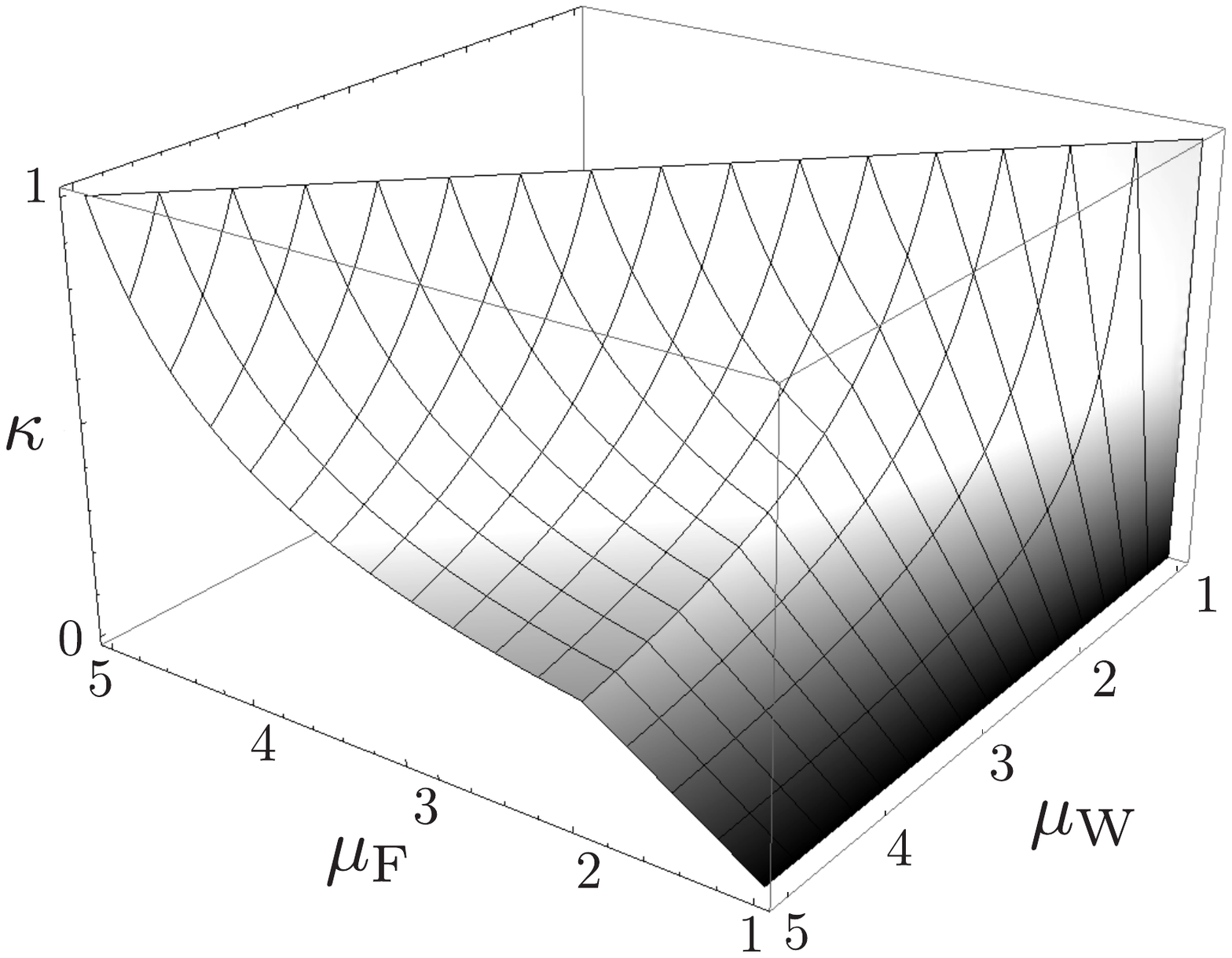}{The demand index $\kappa$ as a function (\ref{kappais}) of $\muf$ and $\muw$.}
{0.45}{189 87 668 469}

More generally, a function $(A-1)/(A-\delta)$ 
with $A>1$ has the required property, but by choosing $A=2$ we obtain
$d\kappa/d\delta|_{\delta=1}=1$, so that $\kappa \simeq \delta$ for $\delta\rightarrow 1$.
This proximity of $\kappa$ and $\delta$ is desirable to some extent
as the data often shows $\delta$ in the range 0.5 to 1.

In summary, the superstatistics framework predicts
that the Pareto indices $\muw$ and $\muf$
determines the Demand Index $\kappa$ as follows:
\begin{equation}
\kappa=
\begin{cases}
\displaystyle
\frac1{\muw-\muf+1} & \mbox{for } 2<\muf;\\[15pt]
\displaystyle \frac{\muf-1}{\muw-1} &\mbox{for } 1<\muf<2.
\end{cases}
\label{kappais}
\end{equation}
This relation is illustrated in Fig.\ref{fig:kappa_mumu}.

\section{Empirical Facts}

\subsection{Database and the Analysis}

In calculating the productivity $c$ by Eq.(\ref{yoverl}) from data,
we calculate
the value added $Y$ by the method put forward by the Bank of Japan
\citep{soumaikeda}, which is the most common method used in Japan.
As for the number of workers $L$, we use the average of the 
value of that year and that of the past year, as each are defined
to be the value at the end of the year.

\figp{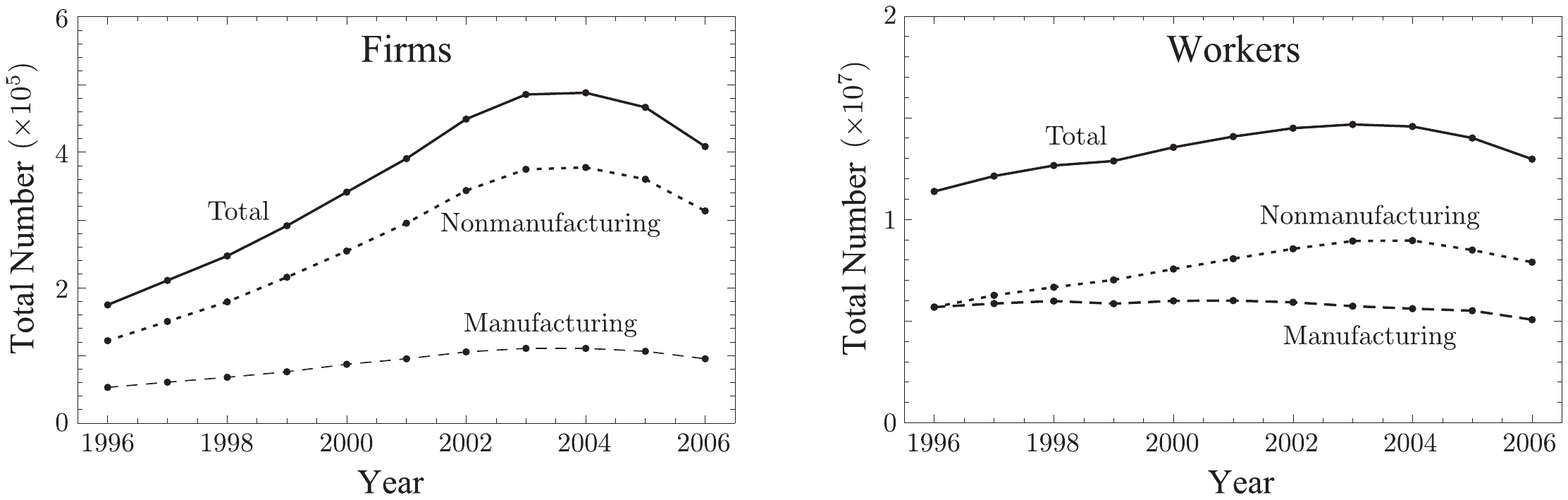}{Total number of firms (left) and workers (right) in our
database, both for manufacturing sector (dashed lines), nonmanufacturing 
sector (dotted lines) and both sectors (solid lines).}
{0.9}{74 205 720 392}

For comprehensive, high-accuracy study, we made a database from two sources:
\begin{description}
\item[Nikkei-NEEDS]
Nikkei Economic Electronic Databank System (NEEDS) database 
is a commercial product available from \citet{nikkei}
and contains financial data of all the listed firms in Japan.
This is a well-established and representative database, 
widely used for various purposes from research to practical business applications.
We have extracted data from their 2007 CD-ROM version,
which contains two to three thousand firms and five to six million workers.
\item[CRD]
Credit Risk Database \citep{crd} is the first and only database for small-to-medium
firms in Japan. It started collecting data from both
banks and credit guarantee corporations \citep{zenshinhoren} since 2001.
The latter, however, does not contain enough database entry necessary for 
the Bank-of-Japan method and thus are omitted from our database.
\end{description}
By combining these two database and removing any overlap, we have
obtained a unified database that covers a wide range of firms in Japan.\footnote{The 
same database was used for productivity analysis by \citet{ikedasouma}.}
The total numbers of the firms and workers covered in this unified database is
plotted in Fig.\ref{fig:nfirmsworkers}.

Unfortunately, CRD data covers only from 1996 to 2006, and does no
go far in the past, unlike Nikkei-NEEDS. This limits our unified 
database to the same short period. We, however, consider that 
it is important to include small-to-medium firms to our analysis so that
we have a view of the whole spectrum of firms and workers in Japan.
Thus we have decided to create and analyse this database. 
In coming years, we trust that CRD will keep collecting data.
So, we may extend our analysis to future, if not past, by continuing and expanding
what we have started here in this paper.

\def\pb{p^{\rm (GB2)}(c)}
\def\xx{\biggl(\frac{c}{c_1}\biggr)}
\def\xxi{\biggl(\frac{c_1}{c}\biggr)}
We need a model distribution to fit the data and
extract the value of the Pareto indices.
It has to have several properties:
(i) It has be defined in the whole range, $c\in [0,\infty]$;
(ii) it has to have power law (\ref{pfasym}) and (\ref{mue}) for large $c$;
and (iii) it has to be able to describe the data to some accuracy
over the whole region.
The power law manifest itself as the
straight line in the log-log rank-size plots of the data 
(see Fig.\ref{fig:2004cdf}.\footnote{Firms with extremely high value of
productivity are removed from this plot, as they often report
{\it one} worker, which, in view of their huge income, cannot be a good representation of
their manpower.})
Since its gradient is the Pareto index, its value 
is estimated fitting the straight section of the
rank-size plot with a straight line. Although this can be 
done easily and is intuitive, it has several pitfalls:
Often, the definition of ``the straight section" is ad-hoc.
Slight change of it can bring nonnegligible change in the value of the Pareto index.
Even if a good one can be found for a particular year,
it might not work for other years of the same database,
which make comparison of different years meaningless.

The ``Generalized Beta Distribution of the Second Kind'' (GB2) \citep{actuarial}
satisfies the property (i)-(iii) and yet manageable.
It is defined by the following pdf;
\begin{align}
\pb
&=\frac{q}{B(\mu/q,\nu/q)}
\frac1c \xx^{\nu}\left[\,1+\xx^q\right]^{-(\mu+\nu)/q},
\label{bb2}
\end{align}
where the four parameters satisfy constraints $\mu,\nu,q,c_1>0$.
Since for large $c$;
\begin{equation}
\pb\simeq \xx^{-\mu-1},
\end{equation}
the parameter $\mu$ is the Pareto index. 
Incidentally, its cumulative distribution functions (cdf) is the following:
\begin{equation}
P^{\rm (GB2)}_>(c)
:=\int_c^\infty \pb \,dc.
=\frac{B(z,\mu/q,\nu/q)}{B(\mu/q,\nu/q)},
\quad
z=\left[\,1+\xx^q\right]^{-1},
\end{equation}
where
$B(z,s,t)$ is the incomplete Beta function with $B(1,s,t)=B(s,t)$.
(Detailed study of small-to-medium productivity was done by \citet{soumaikeda}
using this GB2 distribution.)

\subsection{Manufacturing firms}

\figp{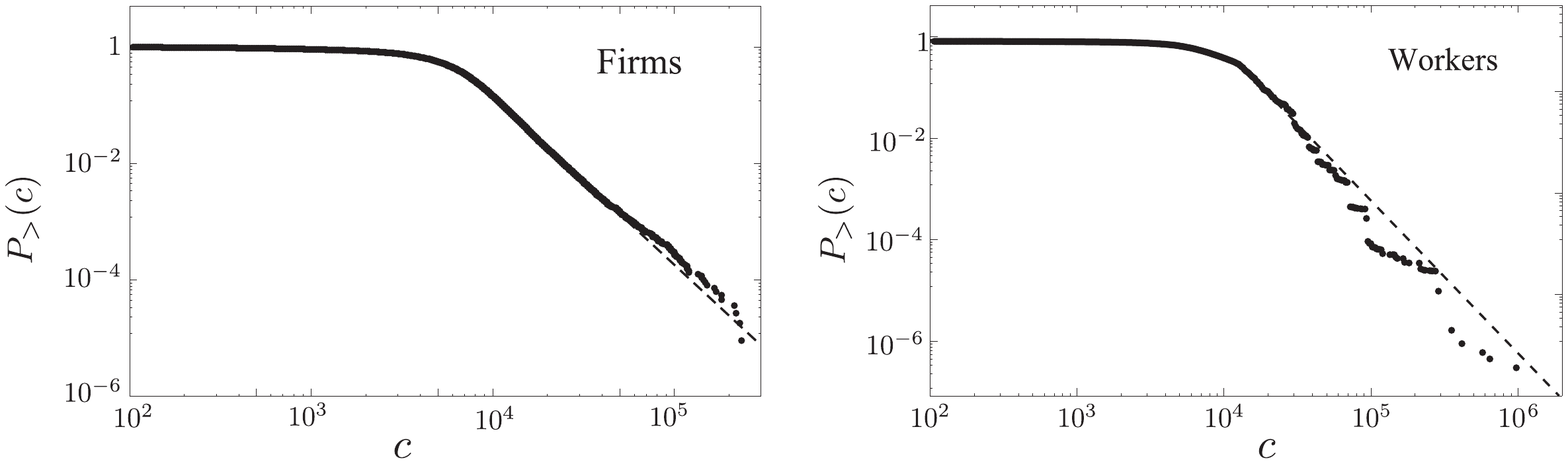}{The rank-size plot of the productivity data (dots)
and the best-fit cdf (dashed line)  for firms (left) and workers (right) in 2004.}
{0.9}{36 174 714 372}
\figp{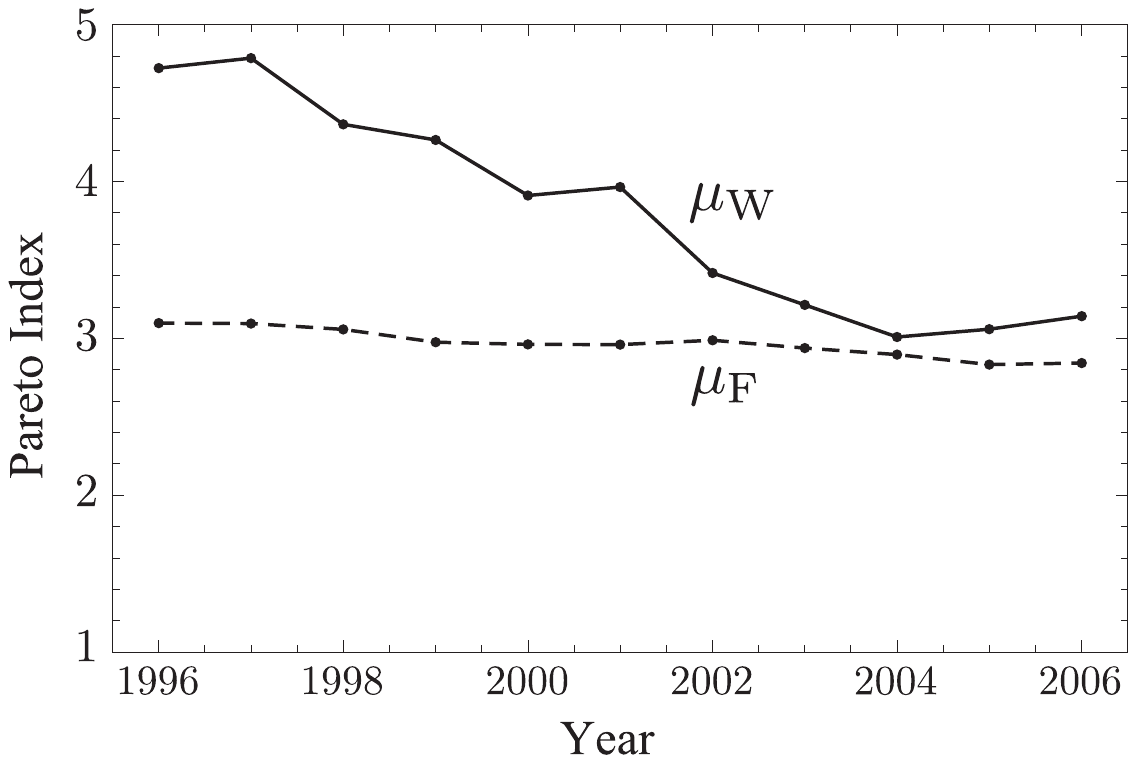}{The Pareto indices $\muf$ and $\muw$ for
the manufacturing sector.}{0.5}{248 181 570 397}

The rank-size plots of the productivity of the manufacturing sector in 2004 
is given in Fig.\ref{fig:2004cdf} by dots,
together with the best-fit cdf
obtained by the maximum likelihood method. 
In these log-log plots, we see that the actual distributions
of the data are close to straight lines for large $c$, which 
implies that it obeys the power law ({\it i.e.}, the Pareto law)
as we have discussed above.
The best-fit cdf (dashed lines) indeed represents the data to good accuracy.
The situation is quite similar in all other years.

The values of the Pareto indices 
$\muf$ and $\muw$ thus obtained are
plotted in Fig.\ref{fig:mu_manufacturing}.
Substituting these values to Eq.(\ref{kappais}), we have obtained the
value of the demand index $\kappa$ joined by solid lines in Fig.\ref{fig:kappa_keiki}.
We see here that the demand is slowing rising during this
period, which is in agreement with general observations in Japan.
Plotted in Fig.\ref{fig:kappa_keiki} with dashed lines is the Nikkei
Business Index (NBI), which is a major business index in Japan \citep{nbi}.
We observe here that their correlation is good to some extent,
which is consistent with the fact that our demand index $\kappa$ provides a measure of demand.

\figp{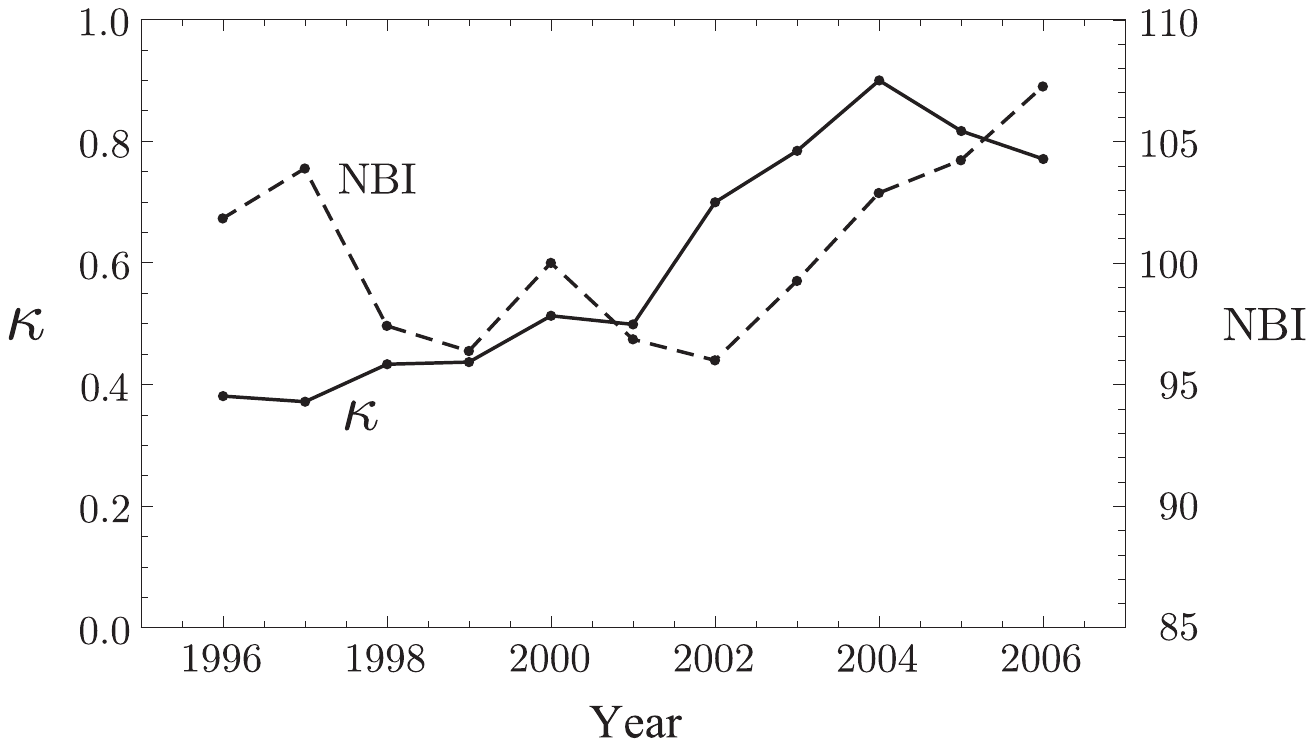}{The demand index $\kappa$ for
the manufacturing sector (solid lines) and
the Nikkei Business Index (dashed lines).}{0.6}{230 182 603 393}

\subsection{Nonmanufacturing (service) firms}
In the nonmanufacturing sector,
the same analysis leads to the result plotted in Fig.\ref{fig:mu_nonmanufacturing}.
It is quite notable that it is completely different from the manufacturing sector:
The Pareto index $\muf$ is {\it larger} than $\muw$.
Since the larger Pareto index means that the pdf is
damped highly for large $c$, this means that
the higher productivity firms, more workers are employed.
This is not allowed under the ordinary Boltzmann distribution
(\ref{pee}) due to the Boltzmann factor $e^{-\beta c}$.
It is not allowed in the superstatistics either, 
since it is an weighted average over the Boltzmann 
distribution. 
Therefore, this behaviour of the nonmanufacturing sector
calls for extension of the theoretical framework.

\figp{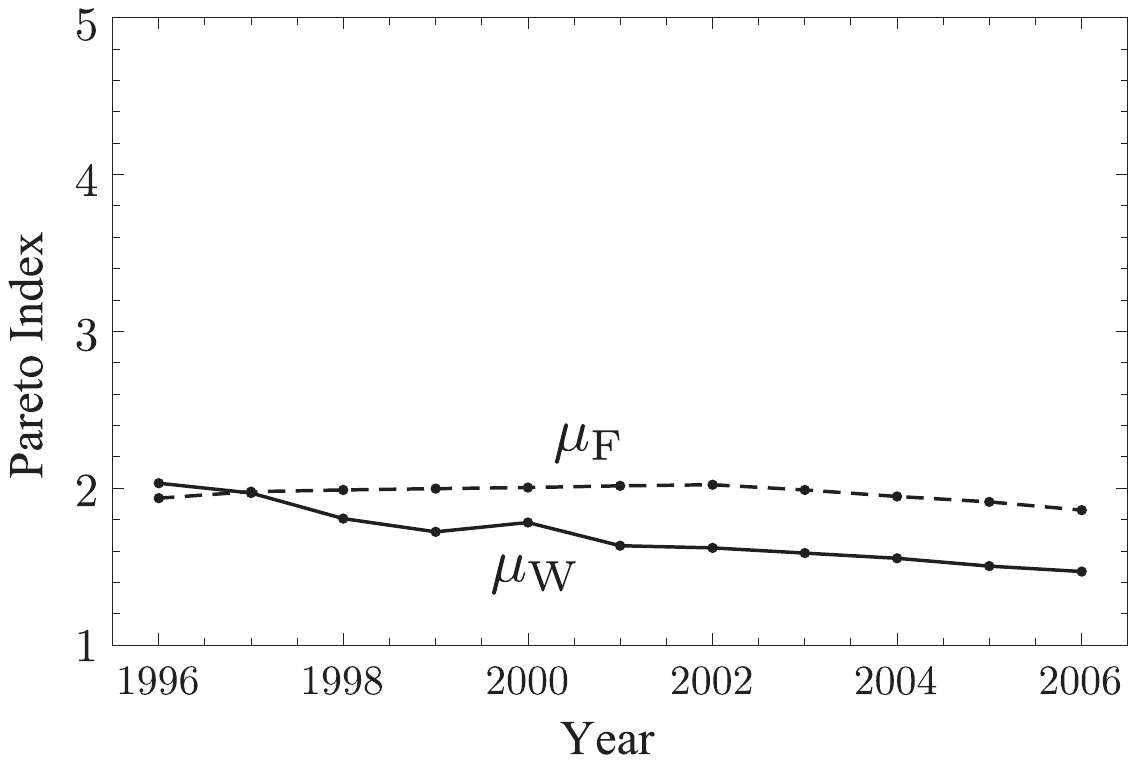}{The Pareto indices $\muf$ and $\muw$ for
the nonmanufacturing sector.}{0.5}{248 179 570 393}

\section{Productivity of Sectors}
Productivity of sectors are of interest.
Our database, following Nikkei NEEDS, contains 26 sectors,
among which 15 are manufacturing sectors and 12 nonmanufacturing sectors.
Their productivity distributions from 1996 to 2006 is plotted 
in Fig.\ref{fig:sectorproductivity_distribution}. Evident in this plot is
that the productivity distribution obeys the power-law with the
Pareto index $\mu\simeq 1.6$ every year. Since the number of
data is limited, unlike the firms and workers, fitting with 
GB2 distribution and estimating the value of $\mu$ is not
very illuminating. In other words, the obtained values of $\mu$
would suffer from large statistical errors. It is more so 
if manufacturing sectors
and nonmanufacturing sectors are studied separately.

\figp{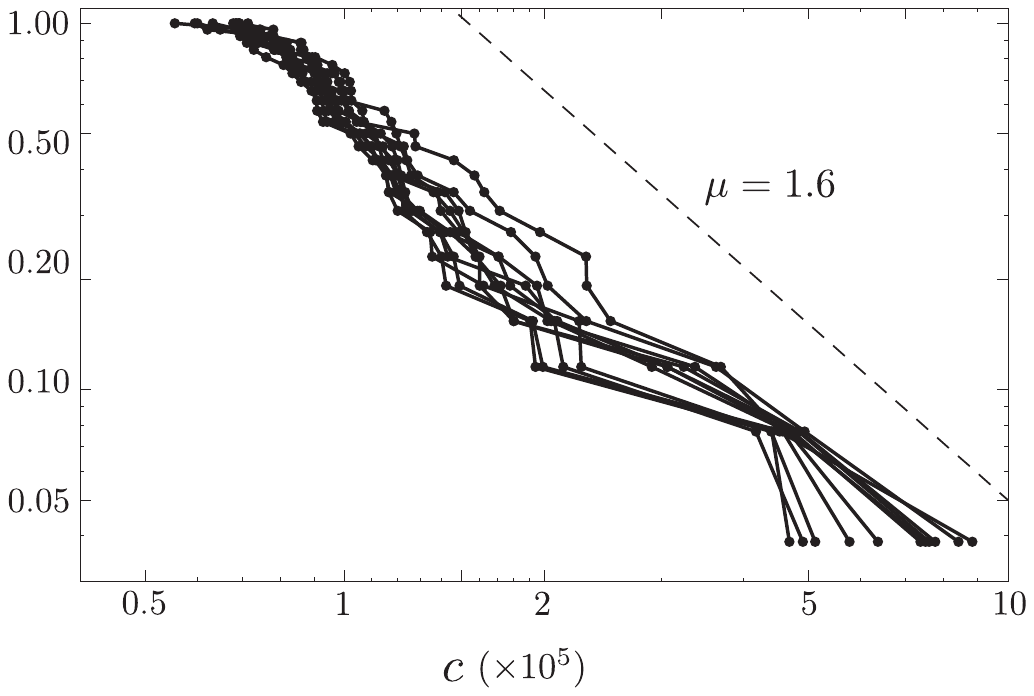}
{Productivity distribution of 26 sectors
from 1996 to 2006. The dashed line is a power-law distribution with
the Pareto index $\mu=1.6$. }
{0.5}{41 183 343 379}

The notable feature of the sector distribution is the
(i) it is approximately Pareto and (ii) the values of
the Pareto index are certainly lower than that of firms every year.
This is in accordance with the idea of applying superstatistics 
framework to firms and sectors, in contrast to workers and firms \citep{ayif1}:
We may now think of a firm (instead of a worker) choosing a sector (instead of a firm) 
for its business activity. Applying the superstatistics to them,
we find again that the Pareto index of the firms are {\it higher}
than that of sectors. This is what we observe in Fig.\ref{fig:sectorproductivity_distribution}.

\section{Conclusion and Discussions}

We have studied the superstatistics theory of productivity
and have proposed the demand index $\kappa$, which determines the
relation between the Pareto indices of the productivity
distributions of firms and workers.
Analysis of the whole spectrum, from small to large, firms
in Japan from 1996 to 2006 is carried out
and manufacturing sector was studied within
the superstatistics framework.

One might argue that what we have been observing is a temporal situation
and eventually, the effect of shocks, including 1999 bubble collapse,
would be averaged out and the productivity starts converging
to a unique value, as predicted by the orthodox equilibrium theory.
In order to study this point, we have examined 
the change of the productivity of sectors over the
years. The result is plotted in Fig.\ref{fig:sectorproductivity_fluctuation}.
Each line represents a sector, solid lines for manufacturing and 
dashed lines for nonmanufacturing.
In this Figure, we see that they are far from arriving at a unique
value. Rather, they keeps fluctuating. Sometimes, the difference
of productivity between the sectors widens. 
This sort of behaviour is typical in physical systems.
The distribution is maintained, while viewed in detail, each atoms
keeps changing its energy and momentum. Such is the
nature of physical equilibrium and so is the economic system.

\figp{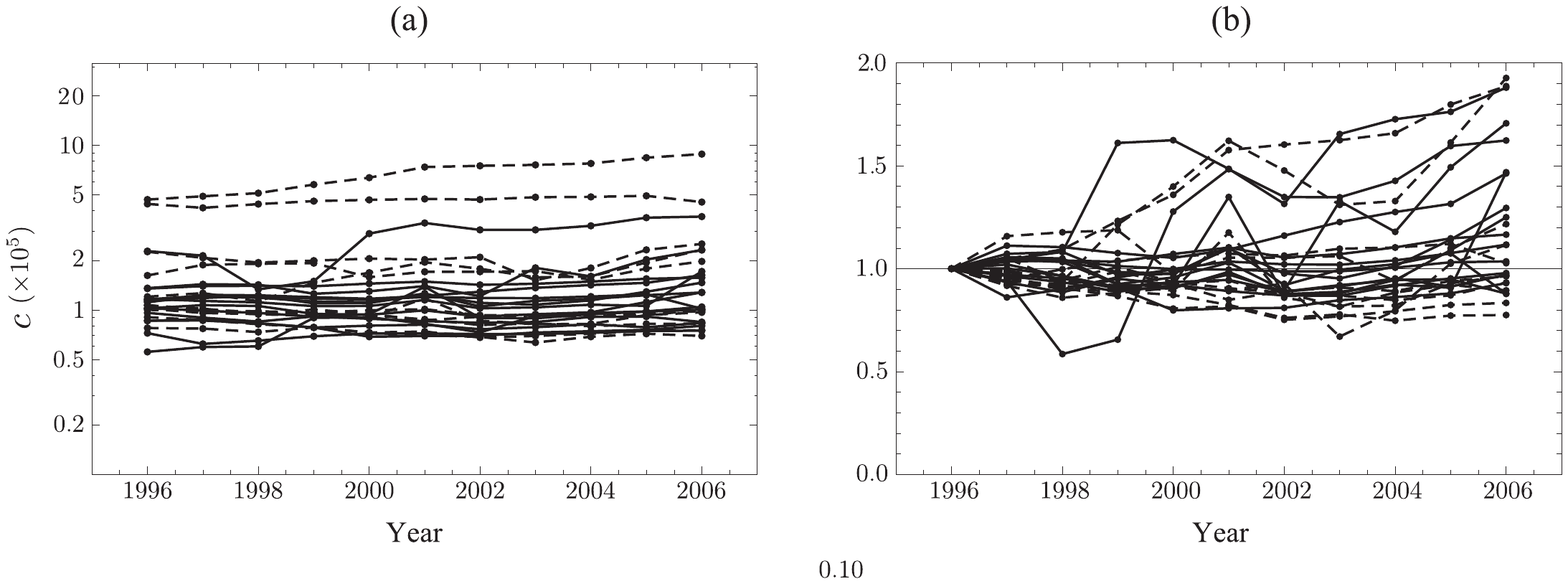}
{Evolution of productivity of each sector. 
sectors. (a) is the actual value, while (b) is the value normalized at 1996.
Solid lines are of manufacturing sectors and dashed lines are of nonmanufacturing sectors.}
{0.9}{50 176 675 406}

We stress here that the distribution (\ref{deltadef}) of the demand $D$ is
unique in the superstatistics framework:
It has to obey this distribution 
if Pareto law hold for both firms and workers and Eq.~(\ref{muwfineq}) is satisfied.
Just what kind of dynamics 
and economic principles underlies the demand distribution (\ref{deltadef})
is unknown. It would be quite interesting and important to 
construct a model which leads to this kind of behaviour.
It is quite possible that such a model could be one of the building blocks useful and necessary for
reconstructing macro-economics.

On the other hand, nonmanufacturing (service) sector
showed peculiar characteristics that has never been
seen before. The Pareto index for the workers were
larger than that for firms.

In the ordinary Boltzmann distribution (\ref{bc}),
the positivity of the temperature guarantees that 
higher the productivity less workers are employed.
Since superstatistics is the weighted average of the Boltzmann
distribution, no matter what the weight function $f_\beta(\beta)$
is, the higher productively means less workers, which is the
reason for $\muf<\muw$.
The fact that the nonmanufacturing sector violates this 
constraint means that the weighted average over the
{\it negative temperature} is required.

Negative temperature is possible for a physical system in nonequilibrium.
One such an example is a laser, where many atoms or electrons 
are in a excited state before the emission.
For the current case, there is at least one economic reason why it is required;
excess of the demand. There are certain limits to the
productivity of a given firm due to many constraints it faces.
But by hiring more people while maintaining the same 
structure, firms can increase its add value, thereby meeting
the increasing demand. Just such cold be happening in the
nonmanufacturing industry.

On the other hand, superstatistics of the negative temperature
has not been developed yet: While it is easy to expand
the weighted average over the negative temperature,
its full consequences are not clear at this stage.
It would therefore be quite interesting to develop this
theory and use it to deal with the nonmanufacturing sector.

%%%%%%%%%%%%%%%%%%%%%%%%%%%%%%%%%%%%%%%%%%%%%%%%%%%%%%%%%%%%%%%
\appendix
\section*{Appendix: Temperature and the Demand}
We first note the following three basic properties (i)--(iii).

\vspace{10pt}
\begin{enumerate}
\item[(i)]
The temperature, $T=1/\beta$ is a monotonically increasing function of the aggregate demand, $D$.
We can prove it using Eq.(\ref{dnbeta}) as follows:
\begin{align}
\frac{dD}{dT}=
-\frac1{T^2}\frac{dD}{d\beta}
=\beta^2\frac{d^2}{d\beta^2}\ln Z(\beta)
=\beta^2\left(\actn{2}-\act^2\right) \ge 0,
\end{align}
where $\actn{n}$ is the $n$-th moment of productivity defined as follows:
\begin{equation}
\actn{n}\equiv\frac1{Z(\beta)}\int_0^\infty c^n \pf \,e^{-\beta c}\,dc.
\label{momdef}
\end{equation}
Note that $\act=D$.
This is a natural result. As the aggregate demand $D$ rises, 
workers move to firms with higher productivity.
It corresponds to the higher temperature due to the weight factor $e^{-\beta c}$.
\item[(ii)]
For $T\rightarrow 0$ ($\beta\rightarrow\infty$),
\begin{equation}
D\rightarrow 0.
\end{equation}
This is evident from the fact that in the same limit the integration
in Eq.(\ref{parti}) is dominated by $c\simeq 0$ due to the factor $e^{-\beta c}$.
\item[(iii)]
For $T\rightarrow \infty$ ($\beta\rightarrow 0$),
\begin{equation}
D\rightarrow \int_0^\infty c\, \pf \,dc\ (=\ac). 
\end{equation}
This can be established based on the property (i) because 
$D=\act\rightarrow \ac$ as $\beta\rightarrow 0$ and 
$Z(0)=1$.
\end{enumerate}

\vspace{10pt}

Let us now study the small $\beta$ (high temperature) properties.
One possible approximation for Eq.(\ref{parti}) is obtained by expanding
the factor $e^{-\beta c}=1-\beta c +\cdots$ and carrying out the
$c$-integration in each term. This leads to the following:
\begin{align}
Z(\beta)&= \int_0^\infty \pf \left(1-\beta c +\frac12(\beta c)^2+\dots 
\right) dc\nonumber\\
&= 1 - \ac\beta + \frac12 \acn{2}\beta^2+ \dots,
\label{notright}
\end{align}
where we have used the normalization condition,
\begin{equation}
\int_0^\infty\pf\, dc =1.
\end{equation}
The result (\ref{notright}) is, however, valid only for $\muf>2$
since $\acn{2}$ is infinite for $\muf \le 2$, 
which is true as we have seen.

The correct expansion for  $1< \muf < 2$ is done in the following way.
We first separate out the first two terms in the expansion of the 
factor $e^{-\beta c}$;
\begin{align}
Z(\beta)&=\int_0^\infty \pf \left(1-\beta c +(e^{-\beta c}-1+\beta c)\right)dc
\nonumber\\
&=1-\ac \beta +Z_2(\beta),\\
Z_2(\beta)&\equiv\int_0^\infty \pf \,g(c)dc
%\nonumber\\ &
=\int_0^\infty \left(-\frac{\partial}{\partial c} \pfc\right) g(c)dc
\nonumber\\
&=\int_0^\infty \pfc \frac{\partial g(c)}{\partial c}dc,
\label{z2good}
\end{align}
where $g(c)=e^{-\beta c}-1+\beta c$ is a monotonically increasing function of $c$ with 
\begin{equation}
g(0)=g'(0)=0.
\label{g00}
\end{equation}
The $c$-integration in Eq.(\ref{z2good}) is dominated by the asymptotic region of $c$
for small $\beta$. Therefore, the leading term in $Z_2(\beta)$ is evaluated by
substituting the asymptotic expression of $\pf$;
\begin{equation}
\pfc \simeq \left(\frac{c}{c_0}\right)^{-\muf},
\end{equation}
into Eq.(\ref{z2good}). We thus arrive at the following:
\begin{align}
Z_2(\beta)&=\int_0^\infty \left(\frac{c}{c_0}\right)^{-\muf} \frac{\partial g(c)}{\partial c}dc
+\cdots
\nonumber\\
&=\muf\Gamma(-\muf)(c_0\beta)^\muf+\cdots.
\end{align}

The case $\muf=2$ can be obtained by taking the limit
$\muf\rightarrow 2+$ in the following expansion valid for $2<\muf<3$:
\begin{equation}
Z(\beta)
= 1 - \ac\beta + \frac12 (\acn{2}-\ac^2)\beta^2
+ \muf\Gamma(-\muf) (c_0 \beta)^\muf+\dots.
\end{equation}
which can be obtained in the manner similar to the above.
The third term is finite for $\muf>2$, but
diverges as $\muf\rightarrow 2+$ as
\begin{equation}
\acn{2}\rightarrow \frac{2c_0^2}{\muf-2}.
\end{equation}
This cancels the divergence of the fourth term in the same limit
and the remaining leading term is as follows:
\begin{equation}
Z(\beta)
= 1 - \ac\beta  - (c_0 \beta)^2 \log(c_0\beta)+\dots.
\end{equation}

In summary, the partition function behaves as follows:
\begin{equation}
Z(\beta)=
\begin{cases}
1- \ac\beta + \frac12 \acn{2}\beta^2+ \dots
& \mbox{for } 2<\muf;\\
1- \ac\beta-(c_0\beta)^2\log(c_0\beta)+\dots
& \mbox{for } \muf=2;\\
1- \ac\beta+ \muf\Gamma(-\muf) (c_0 \beta)^\muf+\dots
& \mbox{for } 1<\muf<2.\\
\end{cases}
\end{equation}
Substituting the above in Eq.(\ref{dnbeta}), we obtain the following:
\begin{equation}
D=
\begin{cases}
\ac-\left(\acn{2}-\ac^2\right)\beta+\dots.
& \mbox{for } 2<\muf;\\
\ac+2c_0^2\beta\log(c_0\beta)+\dots
& \mbox{for } \muf=2;\\
\ac - \muf^2 \Gamma(-\muf) c_0^\muf \beta^{\muf-1} +\dots
& \mbox{for } 1<\muf<2.\\
\end{cases}
\label{hight}
\end{equation}

\newpage
%%%%%%%%%%%%%%%%%%%%%%%%%%%%%%%%%%%%%%%%%%%%%%%%%%%%%%%%%%%%%%%
% bibtex commands
%\bibliographystyle{aer} 
%\bibliography{productivity2}
%%%%%%%%%%%%%%%%%%%%%%%%%%%%%%%%%%%%%%%%%%%%%%%%%%%%%%%%%%%%%%%
% At the VERY last stage of editing:
% 1. comment out the two bibtex commands above.
% 2. insert the final bbl file below.
% 3. edit the included bbl file as necessary, so that the output
%    looks as follows.
% Blouin, M. R. (2003). Equilibrium in a Decentralized Market. 
% Economic Theory 22 (2): 245?262. 
%
% Edwards, S. (1999). How Effective Are Capital Controls? 
% NBER Working Paper 7413. National Bureau of Economic Research, Cambridge, Mass. 
%
% Siebert, H. (1991). The New Economic Landscape in Europe. 
% Oxford: Blackwell. 
% 
% Tharakan, P.K.M., and G. Calfat (1999). Belgium. In M. Brulhart 
% and R. Hine (eds.), Intra-Industry Trade. London: Macmillan. 
%%%%%%%%%%%%%%%%%%%%%%%%%%%%%%%%%%%%%%%%%%%%%%%%%%%%%%%%%%%%%%%
% bbl file
\ifx\undefined\bysame
\newcommand{\bysame}{\leavevmode\hbox to\leftmargin{\hrulefill\,\,}}
\fi

%%%%%%%%%%%%%%%%%%%%%%%%%%%%%%%%%%%%%%%%%%%%%%%%%%%%%%%%%%%%%%%
\end{document}